# A quantum coherent spin in a two-dimensional material at room temperature

Hannah L. Stern†*[1], Carmem M. Gilardoni†[1], Qiushi Gu[1], Simone Eizagirre Barker[1], Oliver Powell[1,2], Xiaoxi Deng[1], Louis Follet[1], Chi Li[4,5], Andrew Ramsay[2], Hark Hoe Tan[3], Igor Aharonovich[4,5] and Mete Atatüre*[1].

[1] Cavendish Laboratory, University of Cambridge, J.J. Thomson Ave., Cambridge, CB3 0HE, United Kingdom

[2] Hitachi Cambridge Laboratory, Hitachi Europe Ltd., CB3 0HE, Cambridge, UK

[3] ARC Centre of Excellence for Transformative Meta-Optical Systems, Department of Electronic Materials Engineering, Research School of Physics and Engineering, The Australian National University, Canberra, ACT, Australia.

[4] School of Mathematical and Physical Sciences, University of Technology Sydney, Ultimo, New South Wales 2007, Australia.

[5] ARC Centre of Excellence for Transformative Meta-Optical Systems, University of Technology Sydney, New South Wales 2007, Australia.

*Corresponding authors. Email: hs536@cam.ac.uk (H.L.S); ma424@cam.ac.uk (M.A)

† These authors contributed equally to this work.

**Abstract:**

Quantum networks and sensing require solid-state spin-photon interfaces that combine single-photon generation and long-lived spin coherence with scalable device integration, ideally at ambient conditions. Despite rapid progress reported across several candidate systems, those possessing quantum coherent single spins at room temperature remain extremely rare. Here, we report quantum coherent control under ambient conditions of a single-photon emitting defect spin in a a two-dimensional material, hexagonal boron nitride. We identify that the carbon-related defect has a spin-triplet electronic ground-state manifold. We demonstrate that the spin coherence is governed predominantly by coupling to only a few proximal nuclei and is prolonged by decoupling protocols. Our results allow for a room-temperature spin qubit coupled to a multi-qubit quantum register or quantum sensor with nanoscale sample proximity.

Scalable spin-photon quantum interfaces require single-photon sources with a level structure that enables optical access to the electronic spins *(1-4)*. These systems are critical for the deployment of applications such as quantum repeaters *(5-8)* and quantum sensors *(9-13)*. Ideally, a spin-photon interface should display long-lived spin coherence with efficient and coherent optical transitions in a scalable material platform without requiring stringent operation conditions such as cryogenic temperature or applied magnetic field. Materials that host atomic defects with well-defined optical and spin transitions garner most attention. Several systems have been studied in detail *(14-21)*, but only a few individually addressable defects in diamond and silicon carbide possess quantum coherent spins at room temperature, albeit with undesirable optical properties *(2,22,23)*. Realising the ideal spin-photon interface requires engineering existing candidates for better performance *(24,25)*, as well as exploring new material systems *(3,26)*.

Two-dimensional layered materials have emerged as a new quantum platform where their natural suitability for large-area growth, deterministic defect creation and hybrid device integration may enable accelerated scalability *(27-35)*. Hexagonal boron nitride (hBN) is a wide bandgap (6 eV) layered material that hosts a plethora of lattice defects that emit across the visible and near infrared spectral regions. The first spin signatures for hBN defects came from ensembles attributed to a boron vacancy ($V_B^-$) with a broad optical emission spectrum centred at ~800 nm *(36-37)*. These defects have been subject to spin initialization and manipulation, albeit exclusively on an ensemble level due to their extremely low optical quantum efficiency *(38-44)*. In contrast, carbon-based defects in the visible spectrum (~600 nm) generate bright, tuneable, single-photon emission of up to 80% into the zero-phonon line (ZPL) at room temperature *(45-50)*. Preliminary reports on these optically isolated defects captured spin signatures via optically detected magnetic resonance (ODMR), but these signatures are only present under finite magnetic field, indicating either a spin-triplet ($S = 1$) with low zero-field splitting or a spin-half ($S = 1/2$) system *(51-53)*.

In this work, we implement room-temperature coherent spin control of individually addressable single-photon emitting defects in hBN. We reveal a spin-triplet ($S = 1$) ground-state spin manifold with 1.96-GHz zero-field splitting using angle-resolved magneto-optical measurements. We observe that the principal symmetry axis of the defect lies in the plane of the hBN layers, indicating a low-symmetry chemical structure. Microwave-based Ramsey interferometry reveals an inhomogeneous dephasing time $T_2^*$ of ~100 ns. Interestingly, the continuously driven spin Rabi coherence time ($T_{Rabi}$) exceeds 1.2 $\mu$s at room temperature with no magnetic field. This drive-prolonged coherence time indicates that the electronic spin can be protected from its reversibly decohering environment, the nuclei. We confirm this via standard dynamical decoupling pulse protocols yielding a spin echo coherence time ($T_{SE}$) of ~200 ns, which exceeds ~1 $\mu$s with ten refocusing pulses. The scaling of the coherence time with the number of decoupling pulses, together with the fine structure of the ODMR signal, suggests hyperfine coupling to only a few inequivalent nitrogen and boron nuclei.

**A ground-state electronic spin triplet**

We investigate multilayer hBN that is grown via metal organic vapour phase epitaxy (MOVPE) using a carbon precursor and ammonia *(54)*, resulting in single-photon emitting and spin-active defects that are related to carbon *(55)*. The flow rate of the carbon precursor determines the defect density in hBN and a 10 $\mu$mol min$^{-1}$ flow rate yields an individually-addressable defect density of ~1 defect per $\mu$m$^2$. The right inset of Fig. 1(a) is a confocal scan image of the hBN device photoluminescence (PL) under 532-nm laser illumination (fig. S1-S2), together with an example second-order intensity-correlation measurement, g$^{(2)}(\tau)$, of a single defect within the scanned area (fig. S3-S4*)*. To identify the ground-state spin resonances we measure continuous wave ODMR, using the protocol presented in the left inset of Fig. 1(a). The main panel of Fig. 1(a) is an example ODMR spectrum for a single defect in the absence of magnetic field and at room temperature. The ODMR signal displays strikingly strong contrast reaching up to ~30% at saturation (fig. S5-6). The spectrum shows two distinct resonances, $v_1$ = 1.89 GHz and $v_2$ = 2.02 GHz. We note that some defects also show the previously reported ODMR resonance when a magnetic field is applied *(52)* (fig. S7-8), indicating the possibility that these two resonances relate to different charge or electronic states of the same defect.

The observation of a zero-field resonance determines a spin multiplicity of > ½ unambiguously. We assign the two resonances to the expected transitions ($\Delta m_s \pm 1$) of a spin triplet with lifted three-fold degeneracy between the spin sublevels, ruling out higher spin configurations (see Section 5 of the Supplementary). An effective-spin Hamiltonian describes the eigenstates of the $S = 1$ system, which in the absence of hyperfine coupling takes the form,

$$H = D(S_z^2 - S(S + 1)/3) + E(S_x^2 - S_y^2) + (g_e \mu_B) \mathbf{B} \cdot \mathbf{S}, \tag{1}$$

where $\mathbf{S}$ is the spin vector with projection operators $S_x$, $S_y$, and $S_z$, $g_e$ is the free electron g-factor, $\mu_B$ is the Bohr magneton and $\mathbf{B}$ is the magnetic field with magnitude $B_0$. The zero-field transitions are characterized by the parameters $D$ and $E$, where $\nu_{1,2} = (D \pm E)/h$ and $h$ is Planck's constant. The ODMR spectrum of Fig. 1(a) yields $D/h = 1.960(10)$ GHz and $E/h = 60(10)$ MHz. To confirm the $S = 1$ assignment we perform vector magnetic field-dependent ODMR measurements. Figure 1(B) displays the evolution of the ODMR transition frequencies as the external magnetic field is applied at $\theta = 0°$ (dark purple circles) and at $\theta = 60°$ (light pink circles), where $\theta$ is the angle between $\mathbf{B}$ and the defect z-axis. We determined the defect z-axis lies in the plane of the hBN layers, with a confidence of 18°, via a series of angular ODMR measurements (fig. S9-12). The purple and pink curves show the simulated transition frequencies between the eigenstates of the $S = 1$ spin Hamiltonian of Eq. 1, with $D/h = 1.959$ GHz and $E/h = 59$ MHz. The same model captures well the appearance of the anticipated $\Delta m_s = 2$ transition in the high off-axis field regime. Interestingly, the ODMR contrast is not quenched up to the highest magnetic field strengths we access, 100-mT magnetic field applied orthogonal to the defect z-axis (fig. S14). Figure 1(c) illustrates the assigned energy-level structure, complete with a ground-state $S = 1$ system with 1.96 GHz zero-field splitting and a metastable spin-singlet state responsible for off-resonance optical spin initialization *(16)*. We measure the ground state resonance across more than 25 defects across multiple devices and obtain a narrow distribution of $D$ and $E$ values: $D/h = 1.971(25)$ GHz and $E/h = 62(10)$ MHz (Fig. 1(d) and (e)). Finally, Fig. 1(f) presents the spin-lattice relaxation ($T_1$) time, where the inset is the protocol used for pulsed ODMR, including a calibrated sequence of laser, microwave and readout pulses for spin initialization, spin control and spin readout, respectively (fig. S17-18). The measured $T_1$ values vary in the range 35-200 μs, which is long compared with the optically excited state lifetime of ~5-6 ns (fig. S4) and that of the previously identified resonance (~ 9 μs (fig. S19)). We conclude that this resonance arises from a ground-state $S = 1$ system.

**Spin coherence and protection under ambient conditions**

Figure 2(a) shows spin Rabi oscillations, where we vary the duration of a resonant microwave pulse while measuring the ODMR contrast. The data (purple circles) are fit to a function of the form $\exp(-\tau/T_{\text{Rabi}})\sin(2\pi\tau\Omega - \varphi)$ (red curve), where $\varphi$ is the phase offset, $\Omega$ is the Rabi drive frequency and $T_{\text{Rabi}}$ is the lifetime of the Rabi oscillations. We confirm $\Omega$ shows a linear dependence on the square root of the microwave power (fig. S20), as expected. Surprisingly, the oscillations persist for over a microsecond ($T_{\text{Rabi}} = 1.250\,(65)\,\mu s$) at a moderate Rabi frequency of $\Omega = 12$ MHz, yielding a $\pi$-pulse fidelity of 0.96(2) and a quality factor (Q = $T_{\text{Rabi}}/T_\pi$) of 25(4) *(56)*. However, as Fig. 2(b) highlights, $T_{\text{Rabi}}$ also displays a notable prolongation as a function of $\Omega$ – a signature behaviour for a spin-rich environment, observed in other III-V materials *(57-58)*. In this regime, the strongly driven spin is effectively decoupled from a slowly evolving nuclear environment akin to motional narrowing *(59)*.

To obtain the bare inhomogeneous dephasing time ($T_2^*$) for the single defect, we perform microwave-based Ramsey interferometry. Figure 2(c) presents Ramsey measurements for 5 values of detuning $\Delta$ between the microwave drive frequency and the $v_1$ transition. Red curves are fit to the data using $\exp(-\tau/T_2^*)\sin(2\pi\tau\Omega_{\text{Ramsey}} - \varphi)$, where $\Omega_{\text{Ramsey}}$ is the frequency of the oscillations arising from $\Delta$. We extract a collective value for $T_2^*$ of 106(12) ns, commensurate with the ~10-MHz linewidth of the unsaturated ODMR spectrum.

Figure 3(a) presents spin coherence measurements using a basic multi-pulse dynamical decoupling protocol, without phase control or pulse corrections, as illustrated above the panel. Spin echo measurements (red circles), comprising a single refocusing pulse, $N_\pi = 1$, yields $T_{\text{SE}} = 228(11)$ ns and $\alpha = 1.81(0.22)$ by fitting to $\exp[-(\tau/T_{\text{SE}})^\alpha]$ (red curve). This value is comparable to the $T_{\text{SE}} \approx 100$ ns determined for $V_B^-$ defect ensembles in hBN under equivalent conditions *(37)(43-44)*. We further prolong the spin coherence time by including additional refocusing pulses. Figure 3(a) displays a range of measurements ($N_\pi = 2, 4, 6, 10$) plotted in orange-to-blue colour-coded data (circles) and the corresponding fits (curves). Figure 3(b) presents the dependence of the protected spin coherence time, $T_{\text{DD}}$, as a function of $N_\pi$, reaching 1.08(4) $\mu$s with ten refocusing pulses. The inset shows the evolution of $\alpha$ values extracted from Fig. 3(a) fits as a function of $N_\pi$, staying above 2 for all values and reaching ~6 for 10 pulses indicating strong protection. In the main panel, the orange curve is a power-law fit yielding a $T_{\text{DD}}$ scaling of $N_\pi^{0.71(4)}$. This scaling exponent falls close to the 0.67 scaling signature expected theoretically for a central electronic spin interacting weakly with a few slowly evolving proximal nuclei *(60-62)*.

**Hyperfine signatures of proximal nuclei**

To understand the nature of the hyperfine coupling and gain insight into the chemical structure of the defect, we investigate the $v_1$ resonance for the presence of spectral fine structure. Figure 4(A) presents a map of Rabi oscillations as a function of microwave detuning, performed at 0 mT and below saturation, where a multi-peak structure is already evident. An integrated line cut of this map between 0 and 100 ns reveals two maxima separated by approximately 10 MHz, as shown in Fig. 4(b). This value is commensurate with the unsaturated ODMR linewidth and the $1/T_2^*$ value. Panels (c) and (d) of Fig. 4 present unsaturated ODMR spectra and Rabi oscillations, respectively, performed with the magnetic field orientated ~60° with respect to the defect $z$-axis. The unsaturated zero-field ODMR spectrum is narrow and asymmetric with a linewidth of order 10 MHz (Fig. 4 and fig. S21). With increasing magnetic field, the ODMR spectrum broadens, and we observe faster dephasing of Rabi oscillations. The purple circles in Fig. 4(e) show the ODMR spectrum for 3-mT magnetic field orthogonal to the $z$-axis ($\theta = 90°$). In this configuration, the ODMR spectrum remains narrow and asymmetric as in the absence of magnetic field. In contrast, Fig. 4(f) presents the ODMR spectrum for 20-mT magnetic field applied parallel to the $z$-axis ($\theta = 0°$). In this case, the ODMR lineshape is symmetric and structured indicating a discrete number of overlapping hyperfine resonances (Supplementary Section 5). The orange shaded curves under the ODMR spectra of panels (e) and (f) are computed ODMR spectra for an $S = 1$ electronic spin coupled to two inequivalent nuclear spins with hyperfine couplings of 13 and 22 MHz (see Supplementary for the model). These nuclear spins can be of different species (e.g. one N and one B), or the same species with different hyperfine couplings. In fact, the computed spectrum of panel (f) corresponds to two $^{14}$N atoms (fig. S22). Strikingly, an $S = 1$ central spin coupled to two inequivalent nuclei is the only model that captures the magnetic field amplitude and orientation dependence of the ODMR spectrum (fig. S23). These results, together with the in-plane symmetry of the defect, provide valuable insight for theoretical efforts aimed at determining the microscopic structure of carbon-based spin-triplet defects in hBN *(63-66)*.

In the low-field regime, the fine structure of the ODMR spectrum is intricately linked to the low-symmetry character of the defect. For an $S = 1$ electronic spin, the non-zero transverse zero-field splitting $E$ gives rise to an avoided crossing at zero magnetic field *(67)*, which may be accompanied by the quenching of a relatively weaker hyperfine interaction. However, if the strength of the hyperfine coupling to neighbouring nuclei is comparable to *E/h,* then it introduces an asymmetric perturbation to the ODMR spectrum, as we observe. The presence of the avoided crossing results effectively in a clock transition with corresponding protection of the electronic spin coherence at zero magnetic field. In the low-field regime, when the magnetic field is applied orthogonal to the defect *z*-axis the clock-transition character of the $v_1$ line is preserved, whereas a magnetic-field applied parallel to the defect *z*-axis brings the system away from the clock-transition point. Therefore, the ODMR spectrum of Fig. 4(e) is reminiscent of the zero-field spectrum of Fig. 4(c), whereas the 3-mT spectrum of Fig. 4(c) is considerably broadened, all commensurate with our identification of a low-symmetry spin-triplet defect.

**Outlook**

An optically addressable *S* = 1 defect with proximal nuclei displaying quantum coherence at room temperature and zero magnetic field in a layered material offers immense promise for quantum technologies. For quantum networks, this system can offer feasible scaling of quantum repeater hardware under ambient operation conditions provided the necessary optical quality is delivered via integrated quantum photonics systems *(68)*. The layered material nature of hBN may facilitate integration into nanostructures for tuning spin and optical properties via application of strain and electric fields without compromising quality. The proximal nuclear spins, coupled to the electronic spin, present an opportunity for the realization of long-lived quantum registers for the electronic spin qubit, a critical element for large-scale quantum architectures *(69-70)*. For quantum sensing, this defect offers a unique and flexible system capable of nanoscale sensing under ambient conditions. The defects we report here are at most 15 nm away from any surface, highlighting their potential as highly proximal nanoscale sensors. We estimate a minimum detectable DC magnetic field of $3\mu T/\sqrt{Hz}$, using $0.77 \cdot (h/g_e \mu_B) \cdot (\Delta v/C\sqrt{R})$, *(71)*, where $\Delta v$ is the linewidth (10 MHz), *C* the contrast (30%) and *R* the measure of brightness ($10^4$ events/s). This estimate is of the same order as that of the well-established nitrogen-vacancy centres in diamond *(71-73)*, with the added advantage of not suffering from defect degradation due to proximity to surface charge noise. Further, the retention of high ODMR contrast under off-axis magnetic field offers flexibility on the dynamic range of vector magnetometry measurements.

# Methods

**Material Preparation**

Multilayer hBN was grown by metal organic vapour phase epitaxy (MOVPE) on sapphire, as described in Chugh *et al. (54)* and Mendelson *et al. (55)*. Briefly, triethyl boron (TEB) and ammonia were used as boron and nitrogen sources with hydrogen used as a carrier gas. Growth was performed at low pressure (85 mBar) and at a temperature of 1350 °C, on sapphire substrates. Isolated defects were generated by modifying the flow rate of TEB during growth, a parameter known to control the incorporation of carbon within the resulting hBN sheets (*55*). The resulting material is ~ 30 nm thick. For confocal photoluminescence measurements, the hBN sheets were transferred to $SiO_2$/Si substrates, using a water-assisted self-delamination process to avoid polymer contamination.

**Experimental Setup**

We perform optical measurements at room temperature under ambient conditions using a home-built confocal microscopy setup. We use a continuous-wave 532-nm laser (Ventus 532, Laser Quantum) that is passed through a 532-nm band-pass filter, onto a scanning mirror and focused on the device using an objective lens with 100X magnification and a numerical aperture of 0.9. We control the excitation power using an acousto-optic modulator (AA Optoelectronics), with the first-order diffracted beam fibre-coupled into the confocal setup. We remove any residual laser light from the collected emission using two 550-nm long-pass filters (Thorlabs FEL550). The remaining light was sent either into an avalanche photodiode (APD) (SPCM-AQRH-14-FC, Excelitas Technologies) for recording photon count traces, or to a CCD-coupled spectrometer (Acton Spectrograph, Princeton Instruments) via single-mode optical fibres (SM450 and SM600) for photoluminescence spectroscopy measurements. We carry out intensity-correlation measurements using a Hanbury Brown and Twiss interferometry setup using a 50:50 fibre beam-splitter (Thorlabs) and a time-to-digital converter (quTAU, qutools) with 81-ps resolution. We used a continuous-wave 405-nm laser (Thorlabs LP405-SF10) via a 405-nm dichroic mirror (Semrock BLP01-405R) to illuminate the sample for charge control.

**Optically Detected Magnetic Resonance**

We perform optically detected magnetic resonance (ODMR) measurements using the confocal setup described above. A copper coil was placed in front of the hBN sample, with optical access through the coil, to provide an out-of-plane oscillating magnetic field. For continuous wave (cw) ODMR a 70-Hz square-wave modulation was applied to the microwave amplitude to detect the change in PL counts as a function of microwave frequency. For measurements under an applied magnetic field, we use a permanent magnet that can be changed in proximity and orientation relative to the device. We perform pulsed ODMR measurements using a pulse streamer (Swabian 8/2)to control a series of switches (Mini-Circuits ZYSWA-2-50DR+) to modulate the laser power, microwave power and signal readout duration.

# References

[1] Awschalom, D. D., Hanson, R., Wrachtrup, J. & Zhou. B. B. Quantum technologies with optically interfaced solid-state spins. *Nat. Photonics,* **12**, 516–527 (2018).

[2] Atatüre, M., Englund, D., Vamivakas, N., Lee, S-Y. & Wrachtrup, J. Material platforms for spin-based photonic quantum technologies. *Nat. Rev. Mater.,* **3**, 38–51 (2018).

[3] Wolfowicz, G. et al. Quantum guidelines for solid-state spin defects. *Nat. Rev. Mater.,* **6**, 906–925 (2021).

[4] Gruber, A., et al. Scanning confocal optical microscopy and magnetic resonance on single defect centres. *Science,* **276**, 2012–2014 (1997).

[5] Kimble, H. J. The quantum internet. *Nature,* **453**, 1023–1030 (2008).

[6] Munro, W. J., Azuma, K., Tamaki, K. & Nemoto, K. Inside quantum repeaters. *IEEE J. Sel. Top. Quantum Electron.,* **21**, 78–90 (2015).

[7] Simon, C. Towards a global quantum network. *Nat. Photon.,* **11**, 678–680 (2017).

[8] Pompili, M., et al. Realisation of a multimode quantum network of remote solid-state qubits. *Science*, **372**, 259-264 (2021).

[9] Maze, J. et al. Nanoscale magnetic sensing with an individual electronic spin in diamond. *Nature,* **455**, 644–647 (2008).

[10] Balasubramanian, G. et al. Nanoscale imaging magnetometry with diamond spins under ambient conditions. *Nature*, **455**, 648–651 (2008).

[11] Taylor, J. et al. High-sensitivity diamond magnetometer with nanoscale resolution. *Nature Phys.,* **4**, 810–816 (2008).

[12] Aslam, N. et al. Nanoscale nuclear magnetic resonance with chemical resolution. *Science*, **357**, 67-71 (2017).

[13] Degen, C. L., Reinhard, F. & Cappellaro, P. Quantum sensing. *Rev. Mod. Phys.,* **89**, 035002 (2017).

[14] Robledo, L. et al. High-fidelity projective read-out of a solid-state spin quantum register. *Nature,* **477**, 574–578 (2011).

[15] Acosta, V. & Hemmer, P. Nitrogen-vacancy centers: physics and applications. *MRS Bull,.* **38**, 127–130 (2013).

[16] Doherty, M. W. et al. The nitrogen-vacancy colour centre in diamond. *Phys. Rep.,* **528**, 1–45 (2013).

[17] Knowles, H. S., Kara, D. M. & Atatüre, M. Observing bulk diamond spin coherence in high-purity nanodiamonds. *Nat. Mater*. **13**, 21-25 (2014).

[18] Christle, D. J., et al. Isolated electron spins in silicon carbide with millisecond coherence times. *Nat. Mater.,* **14**, 160–163 (2015).

[19] Seo, H., et al. Quantum decoherence dynamics of divacancy spins in silicon carbide. *Nat. Commun.,* **7**, 12935 (2016).

[20] Pla, J. J., et al. High-fidelity readout and control of a nuclear spin qubit in silicon. *Nature,* **496**, 334 (2013).

[21] Higginbottom, D. B., et al. Optical observation of single spins in silicon. *Nature,* **607**, 266–270 (2022).

[22] Maurer, P. C., et al. Room-temperature quantum bit memory exceeding one second. *Science,* **336**, 1283–1286 (2012).

[23] Widmann, M., et al. Coherent control of single spins in silicon carbide at room temperature. *Nat. Mater*. **14**, 164-168 (2015).

[24] Aharonovich, I., Greentree, A. & Prawer, S. Diamond photonics. *Nat. Photon.,* **5**, 397–405 (2011).

[25] Lukin, D. M., Guidry, M. A. & Vuckovic, J. Integrated quantum optics with silicon carbide: challenges and prospects. *PRX Quantum*, **1**, 020102 (2020).

[26] Bassett, L. C., Alkauskas, A., Exarhos, A. L. & Fu, K-M. Quantum defects by design. *Nanophotonics*, **8**, 11 (2019).

[27] Geim, A. K. & Grigorieva, I.V. Van der Waals heterostructures. *Nature,* **499**, 419–425 (2013).

[28] Novoselov, K. S., Mishchenko, A., Carvalho, A. & Castro Neto, A. H. 2D materials and van der Waals heterostructures, *Science*, **353**, 6298 (2016).

[29] Kim, S. M,. et al. Synthesis of large-area multilayer hexagonal boron nitride for high material performance. *Nat. Commun.,* **6**, 8662 (2015).

[30] Ma, K. Y., et al. Epitaxial single-crystal hexagonal boron nitride multilayers on Ni (111). *Nature,* **606**, 88–93 (2022).

[31] Xia, F., Wang, H., Xiao, D., Dubey, M. & Ramasubramaniam, A. Two-dimensional material nanophotonics. *Nat. Photon.*, **8**, 899-907 (2014).

[32] Liu, X., et al. Strong light-matter coupling in two-dimensional atomic crystals. *Nat. Photon.*, **9**, 30-34 (2015).

[33] Lin, Z., et al. Defect engineering of two-dimensional transition metal dichalcogenides. *2D Materials*, **3**, 022002 (2016).

[34] Gale, A., et al. Site-specific fabrication of blue quantum emitters in hexagonal boron nitride. *ACS Photon.*, **9**, 2170-2177 (2022).

[35] Elshaari, A. W., et al. Deterministic integration of hBN emitter in silicon nitride photonic waveguide. *Adv. Quantum Technol.*, **4**, 2100032 (2021).

[36] Gottscholl, A., et al. Initialization and read-out of intrinsic spin defects in a van der Waals crystal at room temperature. *Nat. Mater.,* **19**, 540–545 (2020).

[37] Ivády, V., et al. Ab initio theory of the negatively charged boron vacancy qubit in hexagonal boron nitride. *Npj Comput. Mater.,* **6**, 41 (2020).

[38] Gottscholl, A., et al. Room temperature coherent control of spin defects in hexagonal boron nitride. *Science Advances,* **7**, 14 (2021).

[39] Gottscholl, A., et al. Spin defects in hBN as promising temperature, pressure and magnetic field quantum sensors. *Nat. Commun.,* **12**, 4480 (2021).

[40] Baber, S., et al. Excited state spectroscopy of boron vacancy defects in hexagonal boron nitride using time-resolved optically detected magnetic resonance. *Nano Lett.,* **22**, 461–467 (2021).

[41] Mathur, N., et al. Excited-state spin-resonance spectroscopy of V−BB− defect centers in hexagonal boron nitride. *Nat Commun.,* **13**, 3233 (2022).

[42] Liu, W., et al. Coherent dynamics of multi-spin VB− center in hexagonal boron nitride. *Nat. Commun.*, **13**, 5713 (2022).

[43] Haykal, A., et al. Decoherence of $V_B^-$ spin defects in monoisotopic hexagonal boron nitride. *Nat. Commun.,* **13**, 4347 (2022).

[44] Ramsay, A. J., et al. Coherence protection of spin qubits in hexagonal boron nitride. *Nat. Commun.,* **14**, 461 (2023).

[45] Tran, T. T., Bray, K., Ford, M. J., Toth, M. & Aharonovich, I. Quantum emission from hexagonal boron nitride monolayers. *Nat. Nanotechnol.,* **11**, 37–41 (2016).

[46] Cassabois, G., Valvin, P. & Gil, B. Hexagonal boron nitride is an indirect bandgap semiconductor. *Nat. Photon.,* **10**, 262–266 (2016).

[47] Jungwirth, N. R., et al. Temperature dependence of wavelengths selectable zero-phonon emission from single defects in hexagonal-boron nitride. *Nano Lett.,* **16**, 6052–6057 (2016).

[48] Grosso, G., et al. Tunable and high-purity room temperature single-photon emission from atomic defects in hexagonal boron nitride. *Nat. Commun.,* **8**, 705 (2017).

[49] Noh, G., et al. Stark tuning of single-photon emitters in hexagonal boron nitride. *Nano Lett.,* **18**, 4710–4715 (2018).

[50] Proscia, N. V., et al. Near-deterministic activation of room temperature quantum emitters in hexagonal boron nitride. *Optica* 5, 1128–1134 (2018).

[51] Chejanovsky, N., et al. Single-spin resonance in a van der Waals embedded paramagnetic defect. *Nat. Mater.,* **20**, 1079–1084 (2021).

[52] Stern, H. L., et al. Room-temperature optically detected magnetic resonance of single defects in hexagonal boron nitride. *Nat. Commun.,* **13**, 618 (2022).

[53] Guo, N-J., et al. Coherent control of an ultrabright single spin in hexagonal boron nitride at room temperature. *Nat. Commun.*, **14**, 2893 (2023).

[54] Chugh, D., et al. Flow modulation epitaxy of hexagonal boron nitride. *2D Materials*, **5**, 045018 (2018).

[55] Mendelson, N., et al. Identifying carbon as the source of visible single-photon emission from hexagonal boron nitride. *Nat. Mater.*, **20**, 321–328 (2021).

[56] Takeda, K., et al. A fault-tolerant addressable spin qubit in a natural silicon quantum dot. *Science Advances*, **2**, 8 (2016).

[57] Bodey, J. H., et al. Optical spin locking of a solid-state qubit. *npj Quantum Inf.*, **5**, 95 (2019).

[58] Koppens, F. H. L., et al. Driven coherent oscillations of a single electron spin in a quantum dot. *Nature,* **442**, 766–771 (2006).

[59] Bloembergen, N., Purcell, E. M. & Pound, R. V. Relaxation effects in nuclear magnetic resonance. *Phys. Rev.,* **73,** 679–715 (1948).

[60] de Lange, G., Wang, Z. H., Ristè, D., Dobrovitski, V. V. & Hanson, R. Universal dynamic decoupling of a single solid-state spin from a spin bath. *Science*, **330**, 60-63 (2010).

[61] Medford, J., et al. Scaling of dynamic decoupling for spin qubits. *Phys. Rev. Lett.*, **108**, 086802 (2012).

[62] Zarporski, L., et al. Ideal refocusing of an optically active spin qubit under strong hyperfine interactions. *Nat. Nanotechnol.*, **18**, 257–263 (2023).

[63] Babar, R., et al. Quantum sensor in a single layer van der Waals material. *arXiv 2111.09589* (2021).

[64] Jara, C., et al. First-principles identification of single-photon emitters based on carbon clusters in hexagonal boron nitride. *J. Phys. Chem. A*, **125**, 6, 1325-1335 (2021).

[65] Li, K., Smart, T. J. & Ping, Y. Carbon trimer as a 2eV single-photon emitter candidate in hexagonal boron nitride: a first-principles study. *Phys. Rev. Mater.*, **6**, 4 (2022).

[66] Benedek, Z., et al. Symmetric carbon tetramers forming chemically stable spin qubits in hBN. *arXiv, 2303.14110* (2023).

[67] Oniuzhuk, M., et al. Probing the coherence of solid-state qubits at avoided crossings, *PRX Quantum*, **2**, 010311 (2021).

[68] Vogl, T., Lecamwasam, R., Buchler, B. C., Lu, Y. & Lam, P. K. Compact cavity-enhanced single-photon generation with hexagonal boron nitride. *ACS Photonics*, **6**, 8 (2019).

[69] Klimov, P. V., Falk, A. L., Chirstle, D. J., Dobrovitski, V. V. & Awschalom, D. D. Quantum entanglement at ambient conditions in a macroscopic solid-state spin ensemble. *Science Advances*, **1**, 10 (2015).

[70] Gangloff, D. A., et al. Quantum interface of an electron and a nuclear ensemble. *Science*, **364**, 6435, 62-66 (2019).

[71] Dreau, A., et al. Avoiding power broadening in optically detected magnetic resonance of single NV defects for enhanced dc magnetic field sensitivity. *Phys. Rev. B*, **84**, 195204 (2011).

[72] Thiel, L., et al. Probing magnetism in 2D materials at the nanoscale with single-spin microscopy. *Science*, **364**, 973-976 (2019).

[73] Marchiori, E., et al. Nanoscale magnetic field imaging for 2D materials. *Nat. Rev. Phys.,* **4**, 49–60 (2022).

## Acknowledgments

We thank Yuan Ping, Viktor Ivády, Jean-Philippe Tetienne, Leon Zaporski and Martin Hayhurst Appel for helpful discussions.

**Funding:** We acknowledge support from the ERC Advanced Grant PEDESTAL (884745), the Australian Research Council (ARC) through grants CE200100010 and FT220100053, and the Office of Naval Research Global (N62909-22-1-2028). H. L. S. acknowledges a Royal Society fellowship. C.M.G. acknowledges support by the Netherlands Organisation for Scientific Research (NWO 019.221EN.004, Rubicon 2022-1 Science). Q.G. acknowledges financial support by the China Scholarship Council and the Cambridge Commonwealth, European & International Trust. S.E.B. and O.P acknowledge funding from the EPSRC CDT in Nanoscience and Nanotechnology (NanoDTC, Grant No. EP/S022953/1). L.F acknowledges the IDEX international internship scholarship from Paris-Saclay.

**Competing interests:** none declared. **Data and materials availability:** All data needed to evaluate the conclusions in the paper are present in the paper or the supplementary materials.

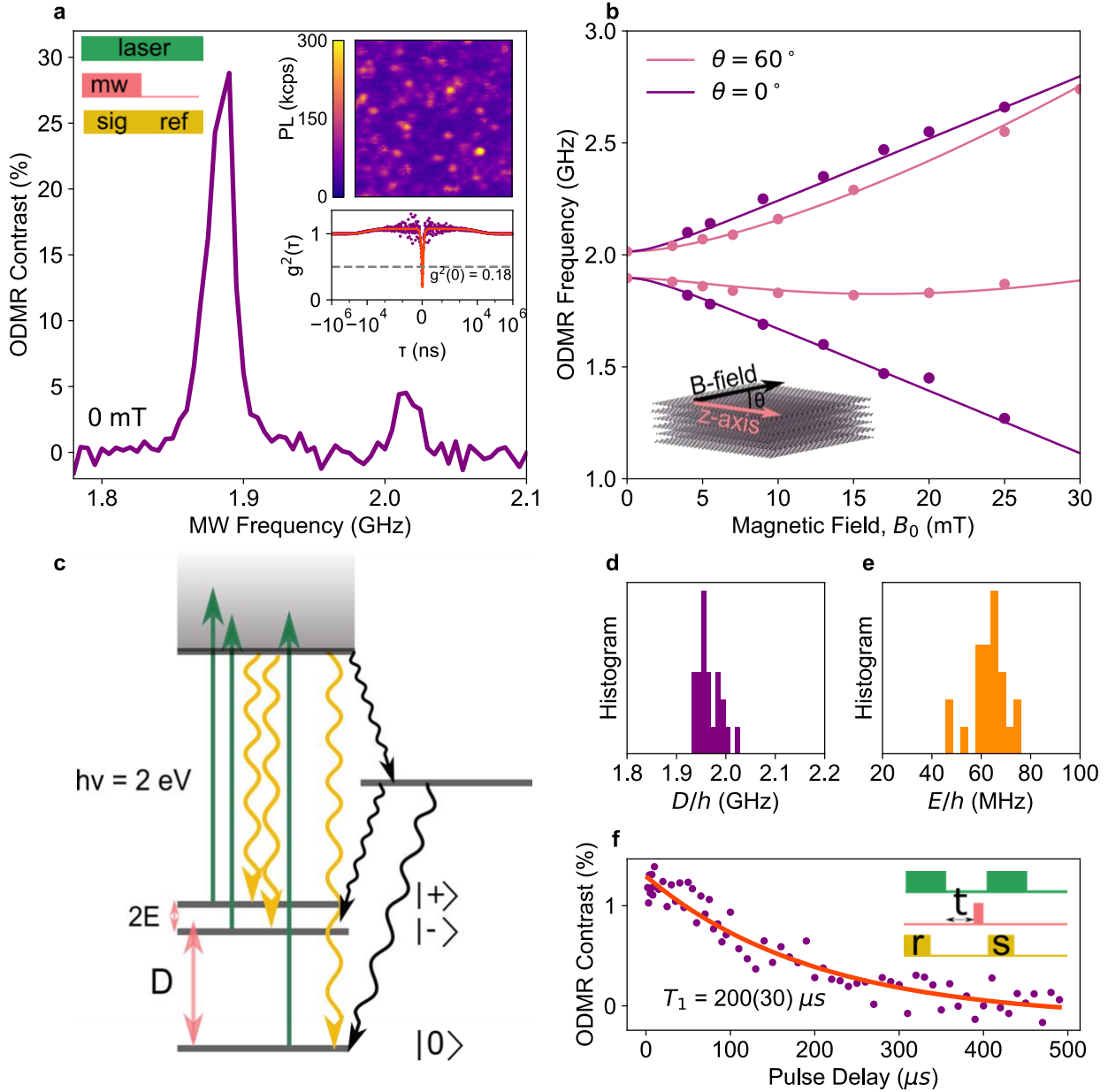

**Fig. 1. A ground-state spin triplet.** (**a**) An ODMR spectrum of a single defect in hBN measured in the absence of magnetic field. The left inset is the measurement sequence. The right inset is a confocal image of the hBN device photoluminescence (PL) under 532-nm laser illumination, together with an example g$^2$($\tau$), second-order intensity-correlation measurement of a single defect. (**b**) ODMR resonance frequencies for a defect, where the magnetic field is applied in the plane of the hBN layers along the defect *z*-axis ($\theta = 0°$) (pink circles), and 60° from the *z*-axis ($\theta = 60°$) (purple circles). Both measurements fit with an *S* = 1 model using Eq. 1 (pink and purple curves, respectively). (**c**) Inferred level structure for the defect, displaying a spin-triplet ground state and spin-singlet metastable state. Green arrows show optical excitation, yellow arrows show PL, black arrows show intersystem crossing and pink arrow shows microwave drive. (**d, e**) statistical

distribution of the $D$ and $E$ zero-field splitting parameters obtained from 25 defects. (**f**) The spin-lattice relaxation ($T_1$) measurement of the ground-state spin, measured in the absence of magnetic field. The inset shows the measurement sequence, not drawn to scale.

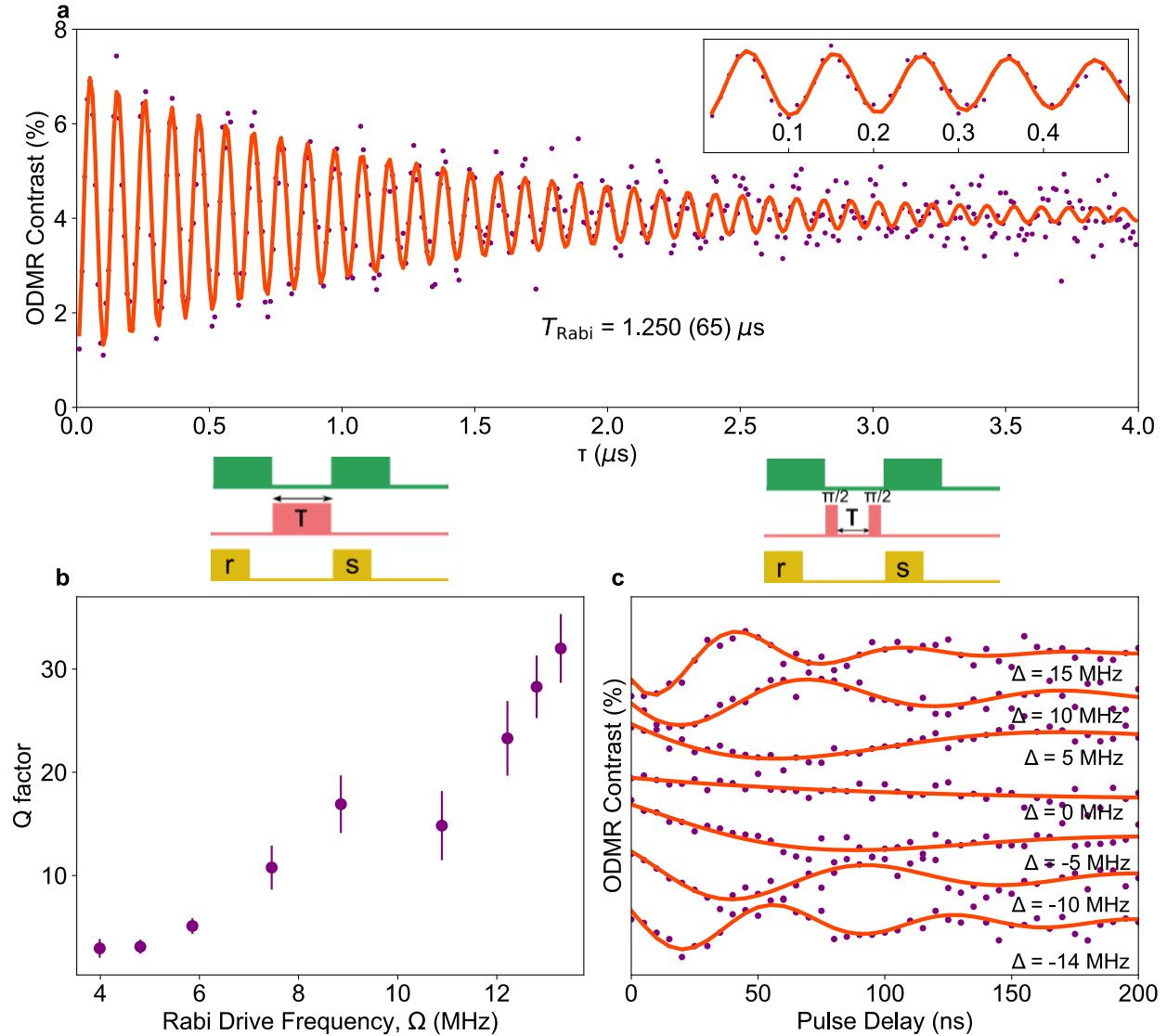

**Fig. 2. Spin coherence and protection.** (**a**) Rabi oscillations (purple circles) for a single hBN defect measured at high microwave power, fit to a function of the form exp($-\tau/T_{\text{Rabi}}$)sin($2\pi\tau\Omega - \varphi$) (orange curve), where $\varphi$ is the phase offset, $\Omega$ is the Rabi frequency and $T_{\text{Rabi}}$ is the decay lifetime of the Rabi oscillations. Inset shows a zoom of the data. (**b**) Q factor of the Rabi oscillations as a function of the Rabi frequency, where Q = $T_{\text{Rabi}}/T_\pi$. The vertical bars indicate one standard deviation. The Rabi measurement sequence is drawn above panel (b) (not drawn to scale), where **r** is reference, **s** is signal and $\tau$ is the pulse duration (**c**) Ramsey measurements for different values of microwave frequency detuning, $\Delta$, from the resonance. Red curves are fit to exp($-\tau/T_2^*$) sin($2\pi\tau\Omega_{\text{Ramsey}} - \varphi$). The diagram above the panel shows the microwave measurement sequence (not drawn to scale), where **r** is reference readout, **s** is signal readout and $\tau$ is the pulse delay. All measurements are at 0 mT.

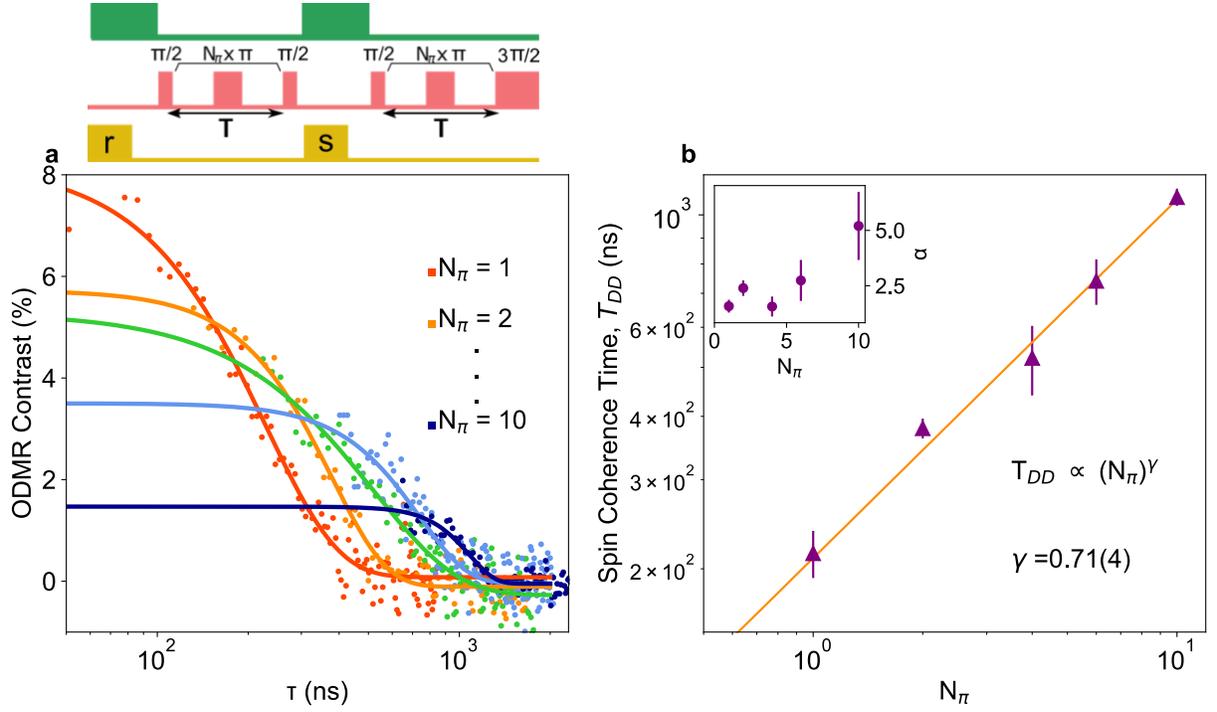

**Fig. 3. Scaling of spin coherence under dynamical decoupling.** (a) Dynamic decoupling measurements with $N_\pi$ refocusing pulses, where each measurement is fit to $\exp[-(\tau/T_{DD})^\alpha]$. The measurement sequence is shown above, where r is reference readout, s is signal readout and $\tau$ is the pulse delay (b) The spin coherence time, $T_{DD}$ (purple triangles) as a function of the number of refocusing pulses, $N_\pi$. The orange curve is fit to a power law, $\sim N_\pi^\gamma$. The vertical bars indicate one standard deviation. All measurements are at 0 mT.

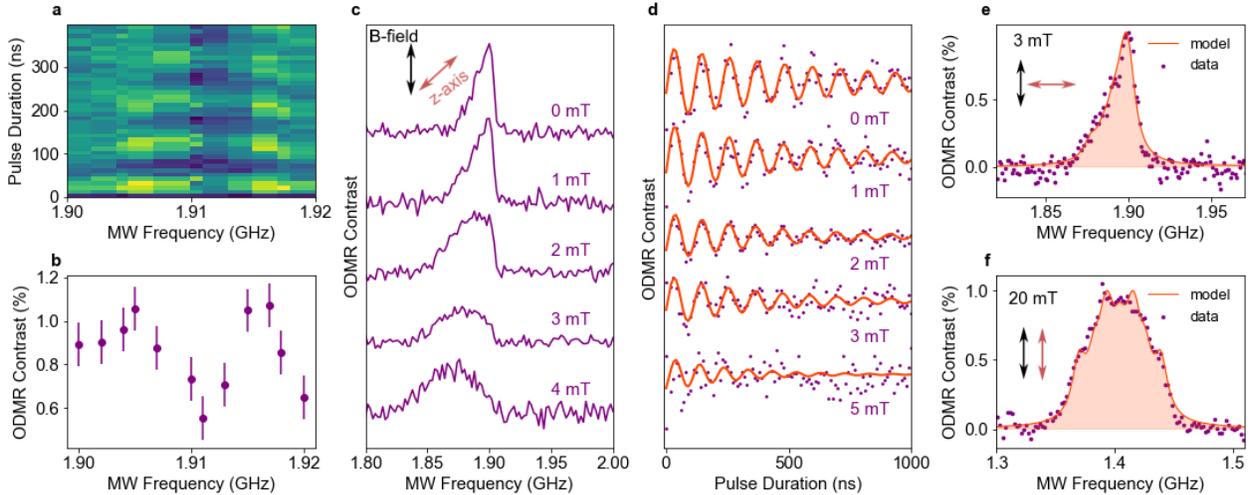

**Fig. 4. Hyperfine signatures in the ODMR spectrum.** (a) Rabi oscillations as a function of microwave frequency, measured at low microwave power. (b) Line cut of panel A, integrated over the first 100 ns. (c and d) ODMR spectra and Rabi oscillations measured between 0 and 5 mT orientated at an arbitrary angle to the defect z-axis. (e) ODMR spectrum measured for 3-mT

magnetic field orthogonal to the defect *z*-axis. (**f**) ODMR spectrum measured for 15-mT magnetic field parallel to the defect *z*-axis. Shaded curves in panels **(e)** and **(f)** are computed spectra.

# Supplementary Information for:

## A quantum coherent spin in a two-dimensional material at room temperature

Hannah L. Stern†*[1], Carmem M. Gilardoni†[1], Qiushi Gu[1], Simone Eizagirre Barker[1], Oliver Powell[1,2], Xiaoxi Deng[1], Louis Follet[1], Chi Li[4], Andrew Ramsay[2], Hoe H. Tan[3], Igor Aharonovich[4] and Mete Atatüre*[1].

† These authors contributed equally.

* Corresponding authors. Email: hs536@cam.ac.uk (H.L.S); ma424@cam.ac.uk (M.A)

# Contents

## 1. Material characterisation

### 1.1 Confocal setup

We use the Qudi software *(1)* to identify ODMR-active defects via automated confocal scans with 100 $\mu$W 532-nm excitation, followed by ODMR measurements at zero magnetic field. A typical confocal scan of a ~5 x 5 $\mu$m region is shown in Fig. S1. In Fig S2. A 100 x 100 $\mu$m region of the sample is shown, where green circles mark the presence of ODMR-active defects.

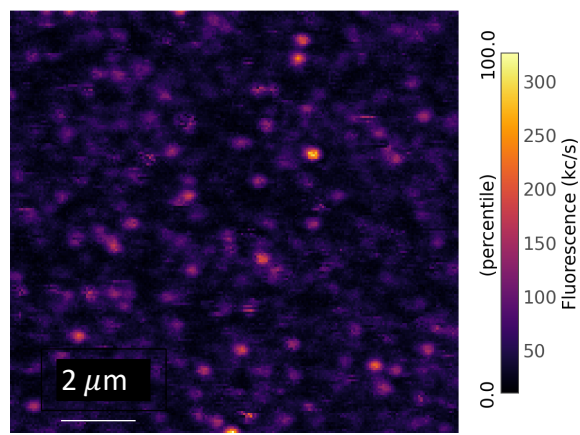

**Figure S1: Confocal scan of hBN.** The photoluminescence count rate for the defects ranges from 50 - 300 x10$^3$ counts per second (c/s) with 100 $\mu$W of 532-nm illumination.

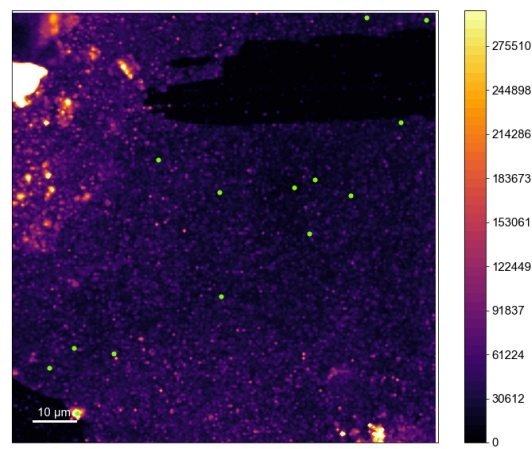

**Figure S2: Confocal scan of a 100 x 100 $\mu$m region of the hBN sample**. The colour bar is in counts per second (c/s) and green circles show the position of ODMR-active defects. Dark regions are the Si/SiO$_2$ substrate, very bright regions are thicker hBN.

### 1.2 Second-order intensity-correlation ($g^{(2)}(\tau)$) measurements.

We perform second-order intensity-correlation ($g^{(2)}(\tau)$) measurements to confirm single photon emission from the ODMR-active defects, via the same method as discussed in *(52)*. Briefly, we performed the $g^{(2)}(\tau)$ measurements by measuring the photon arrival time at two detectors. The recorded arrival times at detector 1 were compared with the arrival times recorded at detector 2, to form an autocorrelation that extends to 1 millisecond separation time. For the data we present, the time stamps from -50 ns to 50 ns are binned in linear-scale time bins, whereas the data at longer time delays are binned in logarithmic-scale bins, to aid visualisation. We used no additional spectral filtering or background correction in the measurement or analysis of the data.

The time-binned data for six ODMR-active defects is shown in Fig. S3 (purple circles), with the corresponding PL spectrum for the same defect (where this was measured) shown in the inset. The data is fit to a linear combination of exponential functions (either two- or three-component fits to functions of the form:

$$f(\tau) = y_0 - a*\exp[-(|x - \tau|\gamma_a)] + b*\exp[-(|x - \tau|\gamma_b)]$$
**(Eq. S1)**

$f(\tau) = y_0 - a*\exp[-(|x-\tau|\gamma_a)] + b*\exp[-(|x-\tau|\gamma_{b1})] + c*\exp[-(|x-\tau|\gamma_{b2})]$,

**(Eq. S2)**

where a, b and c are coefficients, $\gamma_a$ is the antibunching rate and $\gamma_{bi}$ are the bunching rates (i=1,2). From the fit we determine the fitted value of g$^{(2)}$(0) and compare this to the g$^{(2)}$(0) given by the data, and both values are presented in the inset of Fig. S3 and Table S1. We use the antibunching lifetime we determine via the g$^2$(0) measurements to infer the optical lifetime for the defects. The antibunching lifetime is related to both the optical pumping rate and radiative decay rate between the excited optical electronic states and ground states ($\gamma_a = \frac{1}{\tau_a}$ and $\frac{1}{\tau_a} = \frac{1}{k_{opt}} + \frac{1}{k_{rad}}$), therefore the y-intercept of a linear fit to $\gamma_a$ gives an estimation of the optical lifetime *(2)*. The laser power dependent antibunching times for a selection of defects from Fig. S3 are shown in Fig. S4. We determine a radiative lifetime of 5-6 ns.

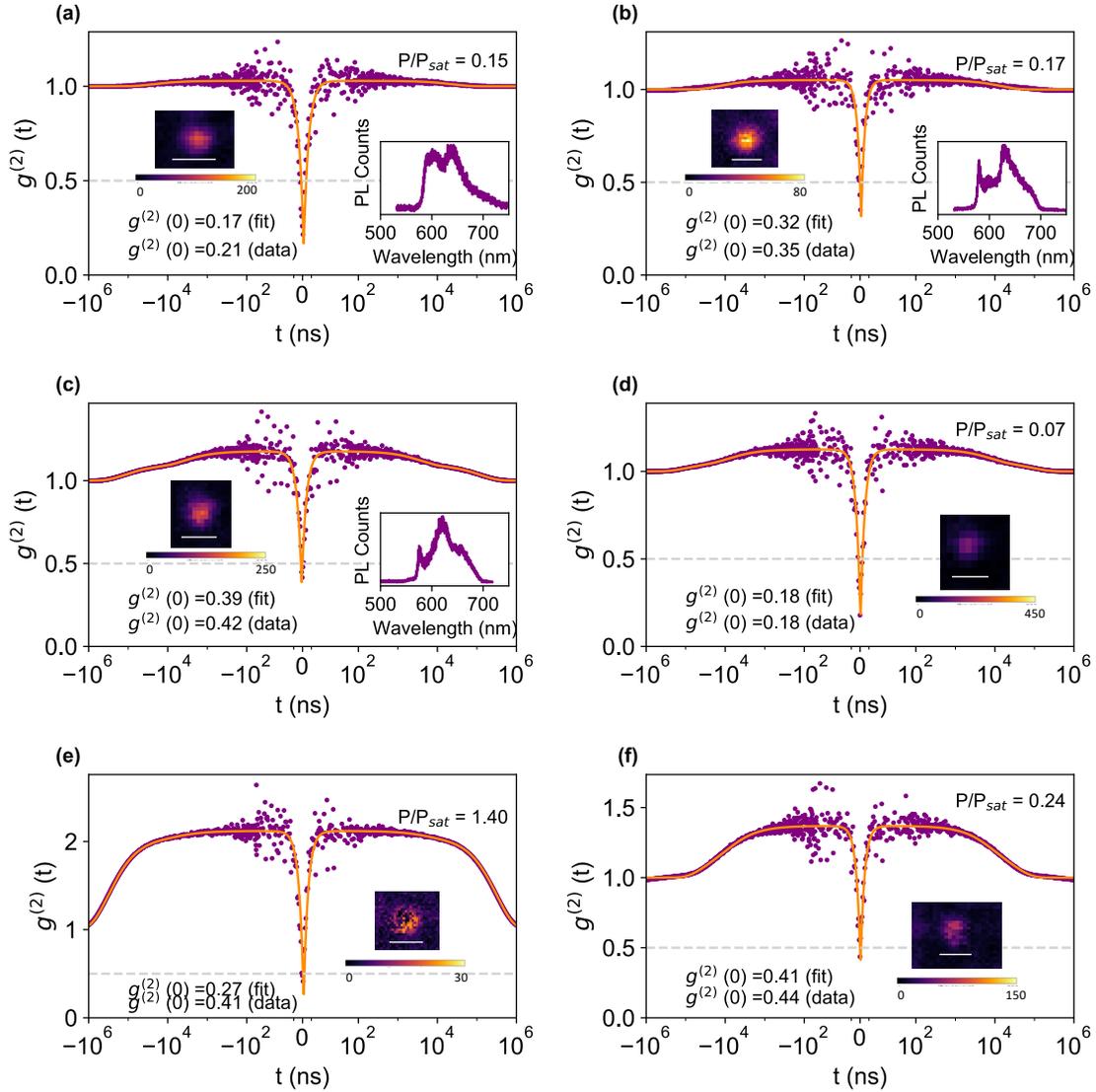

**Figure S3: G$^{(2)}(\tau)$ measurements for six ODMR-active defects we study.** We fit the data (purple circles) to either a bi-exponential or triexponential function given by Eq. 2 and 3 (orange curves). The insets show the integrated photoluminescence spectra (where this was measured) at room temperature using 20-second acquisition time and a 550 nm long pass filter, and confocal scan of the defect (the white scale bar represents 500 nm and z scale is 10$^3$ counts per second (kc/s)). The inset text notes the value of g$^{(2)}$(0) obtained from the data and the fit, as well as the laser power relative to the saturation laser power (*P/P*$_{sat}$) that the measurement was performed at.

**Table S1:** The fit parameters for g²($\tau$) measurements acquired at 50-100 $\mu$W (the laser power used for ODMR measurements) for the defects shown in Fig. S3. Some of the measurements are best fit to two decay timescales (one bunching timescale) but most require three decay timescale (two bunching timescales), as observed for ODMR-active defects in hBN previously *(52)*.

| Defect label (Fig S3) | a | $\gamma_a$ (s⁻¹) | b | $\gamma_{b1}$ (s⁻¹) | c | $\gamma_{b2}$ (s⁻¹) |
|---|---|---|---|---|---|---|
| (a) | 35 | 1.79 ± 0.05 x 10⁸ | 1.2 | 4.08 ± 0.35 x 10⁴ | | |
| (b) | 32 | 2.42 ± 0.08 x 10⁸ | 2.2 | 7.04 ± 1.04 x 10⁴ | | |
| (c) | 66 | 1.93 ± 0.06 x 10⁸ | 5.3 | 2.79 ± 0.36 x 10⁵ | 6.7 | 8.18 ± 1.1 x 10³ |
| (d) | 36 | 2.07 ± 0.05 x 10⁸ | 2.6 | 1.40 ± 0.27 x 10⁵ | 2 | 1.10 ± 0.31 x 10⁴ |
| (e) | 46 | 2.06 ± 0.05 x 10⁸ | 27 | 3.39 ± 0.05 x 10³ | | |
| (f) | 35 | 2.19 ± 0.07 x 10⁸ | 5.4 | 1.72 ± 0.34 x 10⁵ | 7.9 | 3.21 ± 0.43 x 10⁴ |

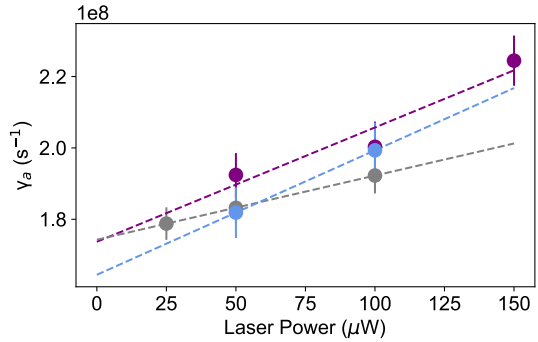

**Fig S4: Estimation of the radiative lifetime.** Antibunching rates as a function of laser power for three defects. The antibunching lifetime is determined by both the optical pumping rate and radiative decay rate between the excited optical electronic states and ground states ($\frac{1}{\tau_a} = \gamma_a$ and $\frac{1}{\tau_a} = \frac{1}{k_{opt}} + \frac{1}{k_{rad}}$), therefore the y-intercept of a linear fit to $\tau_a$ gives an estimation of the optical lifetime *(2)*. In this case, the three defects give $\tau_a$ = 5.75(0.3), 5.72(0.3) and 6.08(0.5) ns.

## 2. ODMR measurements

### 2.1 Determination of microwave and optical saturation conditions.

To determine the saturated contrast and unsaturated linewidth of the ODMR spectra we perform laser power dependent and microwave power dependent measurements of the defects. Figures S5 and S6 show this data for the same defect.

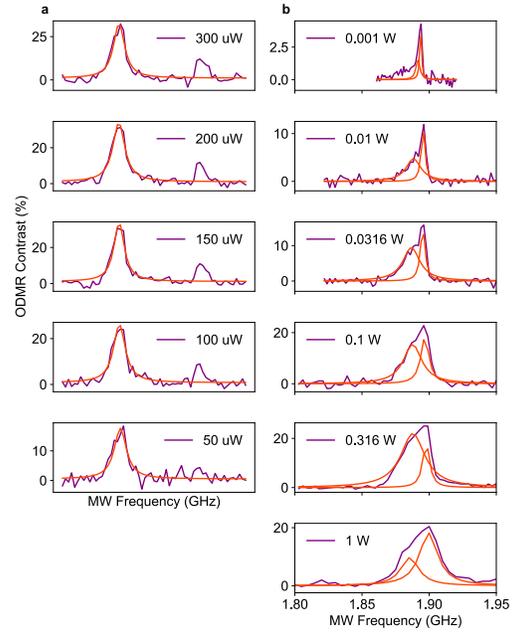

**Figure S5: Microwave and laser power saturation of ODMR at 0 mT.** (a) ODMR spectra measured at different laser powers, at 1W microwave power. (b) Microwave power dependence (0.001-1W) at 100 $\mu$W laser power. The data in (a) (purple lines) is fit to a single Lorentzian (red lines) to obtain the peak contrast. The data in (b) is fit to a double Lorentzian and the fit of the higher frequency Lorentzian is plotted in Figure S6.

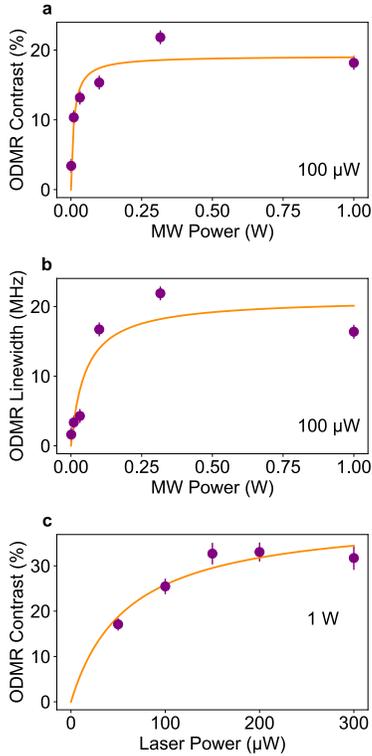

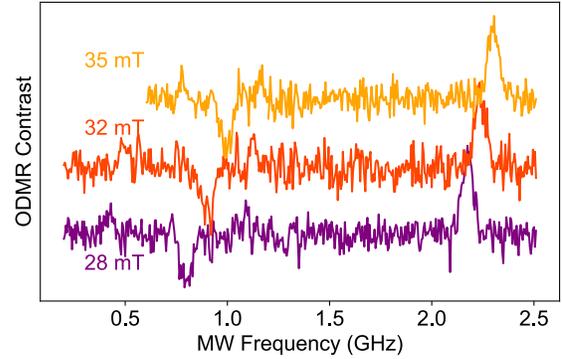

**Figure S7: Observation of multiple ODMR resonances for a single defect**. ODMR spectra for a single defect showing the low frequency, previously identified resonance (~ 800 MHz) and high frequency ground-state resonance studied in this report (~ 2.2-2.4 GHz), at an applied, off-axis magnetic field of 28- 35 mT. This is the defect shown in Fig. S3(c).

**Figure S6: Saturation of ODMR at 0 mT** (a) Microwave power dependence of ODMR contrast and (b) linewidth at 100 $\mu$W microwave power. (c) Laser power dependence of ODMR contrast, at 1W microwave power. All data (purple circles) are fit to C = $C_{sat}$P/P+$P_{sat}$ where C is the ODMR contrast/linewidth, P is the microwave/laser power, $C_{sat}$ is the ODMR contrast/linewidth at saturation and $P_{sat}$ is the saturation laser power.

## 3.2 Observation of the previously reported ODMR resonance

In a previous publication, an ODMR resonance that was not present in the absence of a magnetic field, was identified on the same sample that we study here *(52)*. In *(52)* we assigned this to a $S > ½$ resonance with low zero-field splitting due to the presence of fine structure. We can confirm that some defects display both the ground state spin resonance that we study here and the previously identified resonance (example Figure S7-8). Both resonances cannot be explained by a single spin model, therefore we conclude that the previously identified resonance is likely to be either an excited state ODMR resonance or the resonance of a charge state.

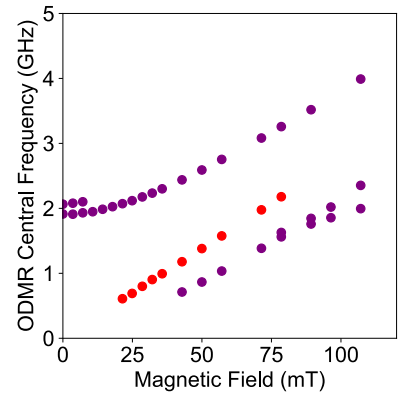

**Figure S8: Field-dependence of multiple ODMR resonances for a single defect**. The ODMR resonance frequencies measured for a defect that shows both resonances (Fig. S7). The purple circles represent the transition energies of the S =1 ground states (including the $\Delta m_s$ = 2 transition > 40 mT) and the red circles represent the ODMR transition energies of the previously identified resonance. This is the defect shown in Fig. S3(c).

## 3.3 Angular-dependent ODMR

To confirm the orientation of the defect's *z*-axis relative to the lab frame we performed a series of angular-dependent ODMR measurements. We use a magnet mounted on an angular mount that allows us to vary the in-plane angle of the magnetic field (ie. the magnet is moved in the plane of the sample).

To present our results, we must define a new set of axes ($x_{lab}$, $y_{lab}$ and $z_{lab}$) in the lab-frame, where $y_{lab}$ is the optical axis and sample lies in the $z_{lab}$ - $x_{lab}$ plane (Fig. S9). The polar angle between **B** and $z_{lab}$ is $\theta'$. In Fig. S10 we present $\upsilon_1$ and $\upsilon_2$ for the same defect that is presented in Fig.1(b) in the main text (and Fig. S12 below), as a function of $\theta'$ ie. the magnet is moved around in the plane of the sample, at 14(1) mT. We find that the ODMR transition energies are very sensitive to $\theta'$. We fit the ODMR transition energies to an $S = 1$ spin Hamiltonian (dark purple curve):

$$H = g\mu_B B(\sin\theta\cos\varphi S_x + \sin\theta\sin\varphi S_y + \cos\varphi S_z) + D(S_z^2 - S(S+1)/3) + E(S_x^2 - S_y^2),$$

**(Eq. 3)**

where $D$ and $E$ and taken from the zero-field ODMR resonance, and $\theta$ and $\varphi$ are defined as the polar and azimuthal angles of the defect axis of highest symmetry (Fig. S9). The fit to the data indicates that $\theta = 24(1)°$ and $\varphi = 0(18)°$ which indicates the axis of highest symmetry is almost in the sample plane (Fig. S11). We confirm this with ODMR measurements taken at $\theta' = 24(5)°$ and varying magnetic field amplitude (Fig. S12). Here, the ODMR transition energies evolve symmetrically with the applied field, consistent with the eigenstates of an $S = 1$ model where **B**∥z (fit shown in Fig. 1 of main text).

We compare the orientation of this defect's z-axis with respect to the lab frame determined in this way to the results we obtain from the field-dependence of the ODMR transition energies for the same defects (Fig. S13 (a)). We find a difference of 6° in the orientation of defect z-axis relative to the lab frame, determined via these two methods. We consider this difference of ~10° is related to the error in $\theta'$ during the rotation, due to the small unavoidable misalignment of the centre of magnet rotation axis relative to the confocal spot. Also, we expect a small change in field strength (~ 1 mT) during the rotation due to this misalignment. Overall, we find the defect z-axis, for all the defects where we performed the angular dependence (six), lies in the plane of the sample.

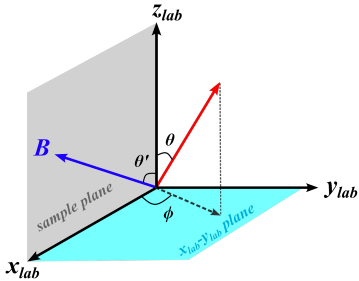

**Figure S9: Definition of the lab frame axes.** Definition of the lab frame coordinates ($x_{lab}$, $y_{lab}$ and $z_{lab}$), where $\theta'$ defines the angle between B (blue vector) and $z_{lab}$, $\theta$ and $\varphi$ are the polar and azimuthal angles of the defect axis of highest symmetry (red vector) in the lab frame.

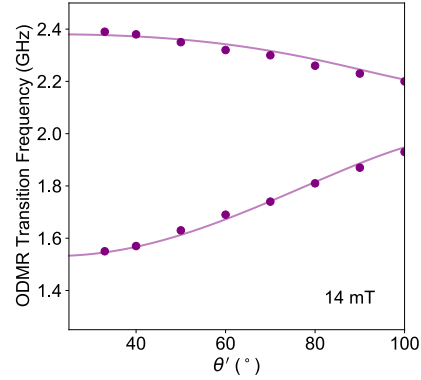

**Figure 10**: **Angle-dependent ODMR.** The ODMR transition energies (circles) and $S = 1$ model (curve) for the in-plane lab-frame angular dependence.

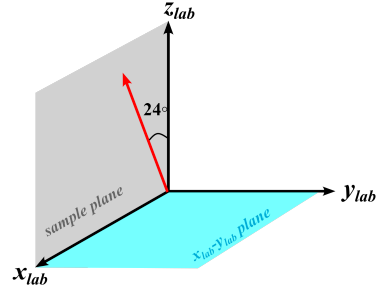

**Figure S11:** Resulting defect symmetry in the lab frame. The axis of highest symmetry (red vector) lies almost in the plane of the sample.

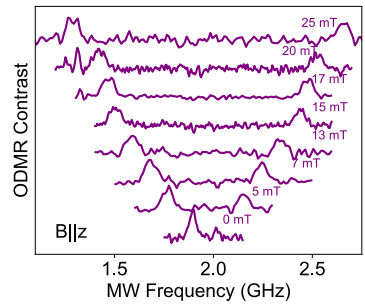

**Figure S12**: **On-axis ODMR spectra.** The ODMR spectra measured where $\theta' = 24(5)°$ for the defect shown in Fig. S10.

## 2.4 Field-dependent ODMR

Once we identify ODMR-active defects at zero magnetic field, we study the field-dependence of the ODMR resonances. We perform these measurements using a calibrated magnet that is fixed to translation stage that gives us access to up to 100-mT applied fields. For most of the defects, we measured up to 40 mT, as this was adequate to get a good fit to the field-dependence. As our sample is mounted vertically, this means the magnetic field is applied along the sample plane, at an arbitrary orientation relative to the *z*-axis of each defect. Figure S14 shows the evolution of the ODMR transition energies for nine different defects in the same sample, as a function of the off-axis **B** (purple circles). The data is fit to the eigenstates of a ground state $S=1$ spin Hamiltonian, where $\theta_0$ and $\varphi_0$ are defined as the polar and azimuthal angles of the B-field in the diagonalised frame of the zero-field tensor (Fig. S13). We determine $\theta_0$ for each emitter and note it in the subplot inset. The value of $\varphi_0$ is indetermined from the field-dependence at a fixed orientation, has little effect on the model and is set to 10°. The simulations provide a good fit to the data, although they do not determine the orientation of the axis of highest symmetry relative to the lab frame, which we determine in Section 3.3.

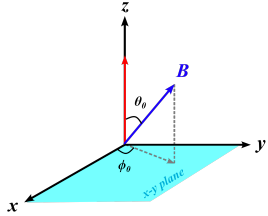

**Figure S13: Axes defined in the diagonalised frame of the zero-field tensor.** The principal axis of **D** with principal value proportional to D (red vector) defines the *z*-axis in this frame. $\theta_0$ is the polar angle of **B** defined as the angle between z and **B** (blue vector), and $\varphi_0$ is the azimuthal angle defined as the angle between x and the projected B-vector in the x-y plane.

Interestingly, for some defects we observe a low-frequency ODMR resonance that we assign to the formally forbidden $\Delta m_s = 2$ transition. This transition is observed when the magnetic field strength is > 40 mT and when **B** is significantly detuned from the orientation of the principal axis. We measure that the ODMR contrast is highest with an on-axis **B**. For an off-axis **B**, the contrast drops as the magnetic field strength increases (Fig. S15), although we note that that contrast is not completely quenched and for some of the defects we measure the contrast remains up to high off-axis -field strengths (~100 mT), unlike for NV centres in diamond *(3)*. The retention of ODMR contrast under high off-axis fields is likely to be related to spin-dependent photodynamics and the nature of the spin mixing in this system, which should be further investigated. Importantly, the sensitivity of the ODMR to off-axis fields may mean this defect is well-suited to magnetometry of systems where detection of fields of tens of mT is required.

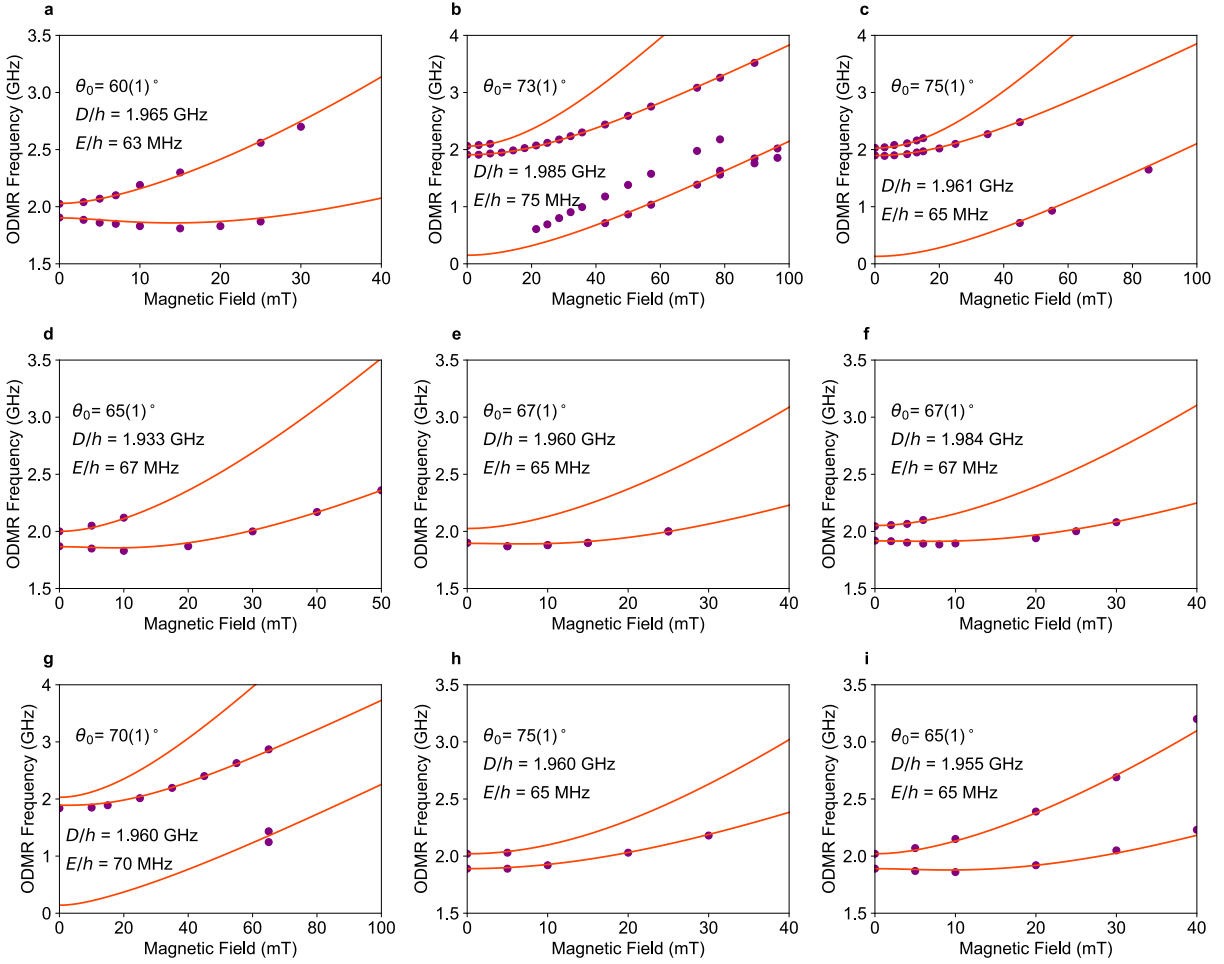

**Figure S14: Magnetic field-magnitude dependence of the ODMR transitions.** Field-dependence of the ODMR resonances (purple circles) for nine different defects, conducted with an in-plane magnetic field. The data is fit to an $S = 1$ model (pink curves). The orientation of the defect $z$-axis relative to **B** ($\theta_0$) is determined from the fitting process and is noted in the inset of each figure.

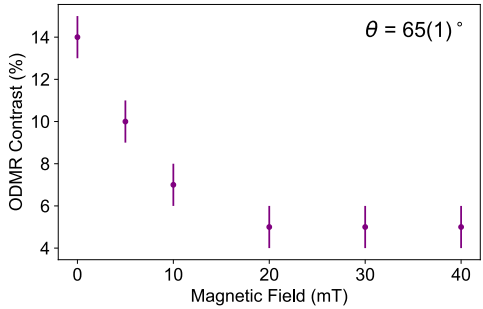

**Figure S15:** Saturated ODMR contrast for a defect with an off-axis B ($\theta = 65(1)°$).

## 2.5 Zero-field ODMR

We identify ODMR-active defects via automated confocal scanning combined with ODMR at zero magnetic field. In Fig. S16 are sample zero-field ODMR-resonances for 20 of the defects we have studied. We fit each resonance of the zero-field ODMR spectrum to obtain the zero-field splitting parameters. Across all defects we study we measure $D/h = 1.970 \pm 0.02$ GHz and $E/h = 62 \pm 7$ MHz. The scanning range was constant between 1.7 and 2.1 GHz. The saturated ODMR contrast varies between the defects (1 - 30%).

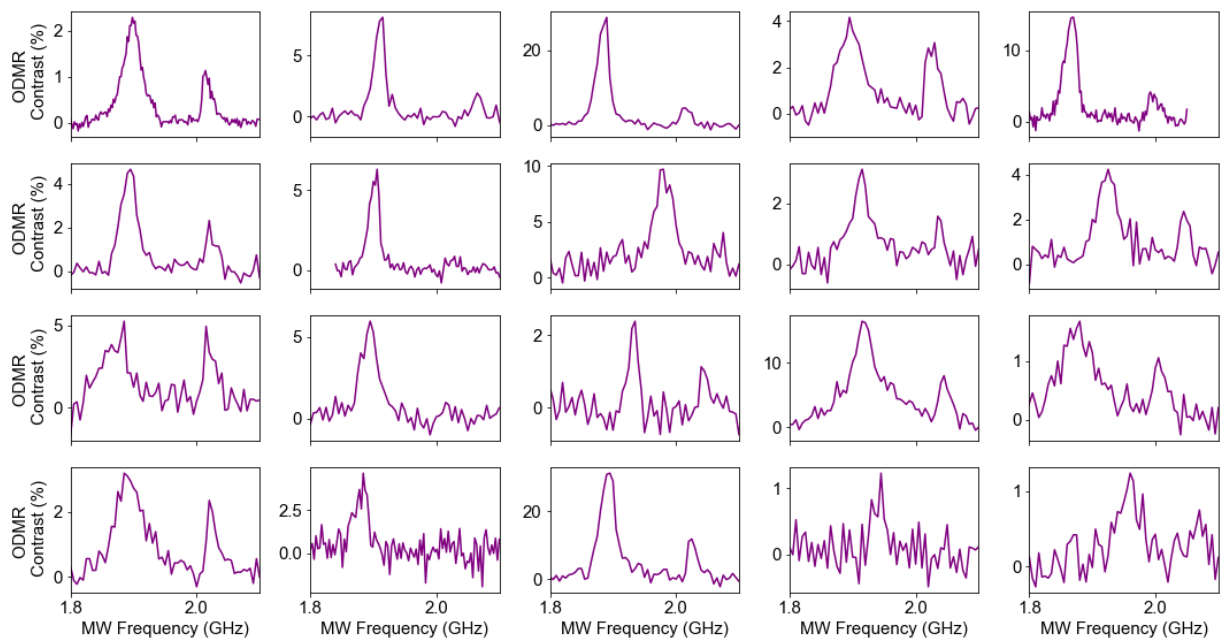

**Figure S16: Zero-field ODMR of single defects.** These measurements have been obtained at high microwave power relative to saturation and 100-200 $\mu$W laser power.

# 3. Coherent microwave control

## 3.1 Readout calibration and pulse sequences

To perform pulsed ODMR measurements we determine the optimum initialisation and read out duration via a calibration measurement. In this measurement we apply a long green pulse followed by a microwave pulse (50 ns) and a second initialisation pulse accompanied by a 500 ns readout pulse. The readout window is iteratively delayed relative to the start of the initialisation pulse. Figure S17 shows the ODMR contrast measured for this measurement as a function of the readout delay. When the contrast returns to zero, the system is fully initialised, which in this case is ~ 80 $\mu$s. We find that most defects require 30-100 $\mu$s initilisation with 100 $\mu$W of 532 nm, consistent with the bunching dynamics (fig. S3 and Table S1).

The optimum readout duration is determined from analysing the signal to noise (accumulated contrast vs. accumulated shot noise) as a function of readout start and stop time (Figure S18). Typically, the optimal readout duration is 20-40 $\mu$s.

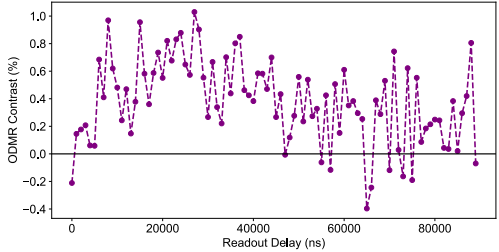

**Figure S17: Spin intialisation.** ODMR contrast (%) as the readout window is scanned over the initialisation laser pulse (readout delay (ns)).

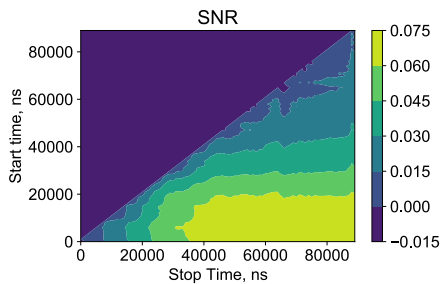

**Figure S18: Readout duration dependence of signal-to-noise.** ODMR signal-to-noise determined as a function of readout start and stop time. The colour bar show signal-to-noise. In this case, the optimal readout duration is 60 $\mu$s.

## 3.2 $T_1$ measurement of the previously studied resonance.

The values for the spin lattice relaxation time that we measure for the ground-state resonance vary from 35 - 200 $\mu$s. This is much longer than $T_1$ measured for single hBN defects previously *(52,53)*. Below we show the $T_1$ measured for the ODMR resonance that we have previously reported in *(52)*. Here, the $T_1$ is ~ 9 $\mu$s, measured at a field strength of 70 mT.

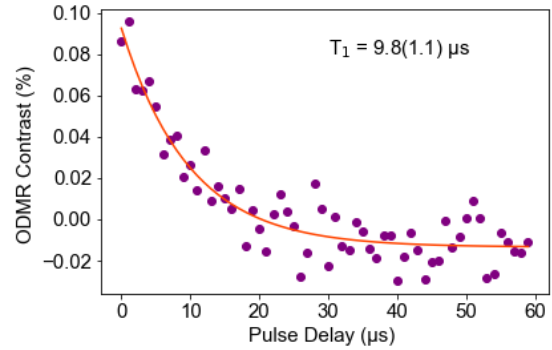

**Figure S19: Spin-lattice relaxation of a separate ODMR resonance.** $T_1$ measurement for the ODMR resonance measured in Stern, Gu, Jarman *et al*. *(52)*.

## 3.3 Rabi measurements

We fit the Rabi oscillations presented in Fig. 2 of the main text to a function of the form,

$$y = A \exp[-(\tau/T_{Rabi})]\sin(2\pi\Omega-\varphi) \quad \text{(Eq. S4)}$$

where $T_{Rabi}$ is the decay of the Rabi envelope, $\Omega$ is the Rabi frequency of each component of the Rabi and $\varphi$ is the phase offset. Below is the dependence of the Rabi frequency on the square root of the microwave power. We observe a linear relation, as expected (Fig. S20).

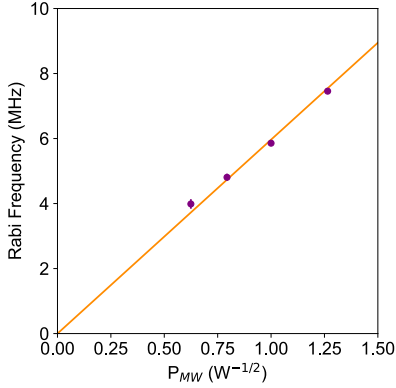

**Figure S20: Rabi frequency microwave power dependence.** Rabi frequency as a function of the square root of the microwave power (W$^{-1/2}$).

### 3.4 Pulsed ODMR lineshape

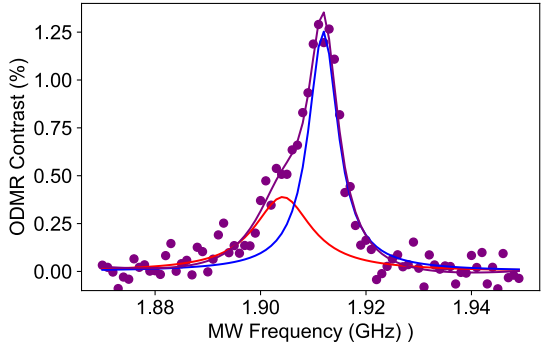

**Figure S21: Unsaturated Pulsed ODMR spectrum.** We fit the data to two Lorentzians, labelled $v_{1b}$ and $v_{1a}$. The peak separation is 10 MHz and respective linewidths are 13 MHz and 6 MHz.

## 4. Spin Hamiltonian

### 4.1 Electron spin Hamiltonian model and $S=3/2$

In the main text we model the magnetic-field dependence of the ODMR spectra based on an effective-spin Hamiltonian composed of zero-field and Zeeman terms. The observation of zero-field ODMR resonances requires a spin-Hamiltonian with multiplicity > 2 (that is, an effective-spin system $S > 1$). We consider both $S = 1$ and $S = 3/2$ possibilities but find that $S = 3/2$ is inconsistent with our data, as discussed below. Larger effective-spin models are highly unlikely since they would require electronic wavefunctions containing more than 4 unpaired electron spins, or eigenstates with large orbital angular momenta.

An $S = 3/2$ system may arise for a wavefunction with three unpaired electron spins. In this case, the exchange interaction between the electrons causes a splitting between two Kramers doublets at zero magnetic field. The zero field ODMR spectrum would show a single zero-field ODMR line at an energy given by $v = (\sqrt{D^2 + 3E^2})/h$. We routinely measure a doublet at zero field, so $S = 3/2$ is unlikely.

In the high-field regime (when the Zeeman interaction is much larger than the zero-field splitting between the levels), an $S = 3/2$ system gives rise to two ODMR-active transitions that are equally spaced and evolve linearly with magnetic field amplitude. Importantly, the two ODMR transitions are equally spaced from the position of a free $S = 1/2$ resonance. This is not what we measure for our defects (see Figure S14). Furthermore, a $S = 3/2$ system cannot explain the low frequency $\Delta m_s = 2$ transition that is observed in our measurements.

Finally, a wavefunction with three unpaired spins may arise in systems with cubic symmetries, where the high symmetry allows an orbital triplet to appear. However, such high symmetry is impossible in a two-dimensional system. Alternatively, the unlikely scenario where the system shows accidental degeneracy between an orbital doublet and an orbital singlet may lead to a wavefunction with three unpaired spins. Nonetheless, in total, the qualitative description of the magnetic-field dependence of the ODMR spectrum of an effective-spin 3/2 system fails to encompass the behavior we measure for the emitters presented in this work, and we discard the possibility that we are dealing with a spin-3/2 system.

### 4.2 Hyperfine models

**Low-field behaviour.**
We can understand the ODMR spectra lineshape in the few-mT regime if we assume that the lineshape of the ODMR transitions is governed by hyperfine interaction with the nuclear-spin bath surrounding each defect.

At zero field, the transverse zero-field splitting parameter $E$ gives rise to an anticrossing related to a clock transition. In this regime, the eigenstates of the system are given by $|0\rangle = |0_z\rangle$, $|\pm\rangle = \frac{1}{\sqrt{2}}(|+1_z\rangle \pm |-1_z\rangle)$. These eigenstates have zero expectation value for the electronic spin projection along any axis ($\langle S_x \rangle = \langle S_y \rangle = \langle S_z \rangle = 0$), leading to a collapse of the hyperfine coupling to neighboring spins. This is consistent with the broadening of the ODMR linewidth upon application of a small magnetic field (Fig. 4 main text), and the dampening of the Rabi oscillations with an applied field (as discussed in the main text). Clock transitions are observed for other calculated spin-1 lattice defects in hBN *(4)*, bulk semiconductor materials *(5)* and molecular qubits *(6)*.

Finally, the asymmetry in the zero-field lineshape arises due to the hyperfine coupling to neighboring nuclei being comparable to the transverse zero-field splitting parameter $E/h$ (60-70 MHz), leading to perturbative mixing between the eigenstates $|+\rangle$ and $|-\rangle$.

**High-field behaviour**

We attempt to model the ODMR spectrum lineshape at an applied field along the defect z-axis by considering the effective-spin Hamiltonian of Eq. 1 of the main text, modified to include the interaction with the nearest neighboring spins

$$H = H_e + \sum_i \vec{S} \cdot A_i \cdot \vec{I}_i + \gamma_{n,i} \vec{B} \cdot \vec{I}_i$$

**(Eq. S5)**

where $\vec{I}_i$ is the nuclear-spin operator associated with the i-th nuclear spin, $A_i$ is the hyperfine coupling tensor describing the coupling between the i-th nuclear spin and the electronic spin and $\gamma_{n,i}$ is the gyromagnetic ratio of the i-th nuclear spin and $H_e$ is the Hamiltonian presented in Eq. 1 of the main text. For simplicity, we consider only Fermi-contact interaction terms. We assume that the only non-zero terms in the hyperfine tensor $A_i$ are $A_{i,xx}$, $A_{i,yy}$ and $A_{i,zz}$.

To model the lineshape of the ODMR spectrum, we calculate the eigenlevels $E_k$ and hyperfine-coupled eigenstates $|\psi_k\rangle$ of the Hamiltonian in Eq. S5, when $\vec{B}_0$ has a magnitude of approximately 20 mT and points in the direction parallel to the z-axis of the defect. When the magnetic field is parallel to the symmetry axis of the defect, the hyperfine coupling terms $A_{i,xx}$, $A_{i,yy}$ act only perturbatively, such that within the secular approximation we only consider $A_{i,zz}$ terms. We assume that each transition at energy $E_{kj} = E_k - E_j$ has a probability given by $P_{kj} = \left|\langle\psi_k|\gamma_e \vec{B}_1 \cdot \vec{S}|\psi_j\rangle\right|^2$, where $\vec{B}_1$ is an oscillating field pointing in the out-of-plane direction with respect to the hBN layers. We assume that the lineshape associated with each individual transition is Lorentzian, with a linewidth of approximately 7 MHz in line with the zero-field pODMR linewidth (fig. S19).

Finally, we assume that nuclear-spin sublevels are all equally occupied. Using these assumptions, the simplest model that qualitatively describes the lineshape we observe when a magnetic field is applied parallel to the symmetry axis of the defect corresponds to a central electronic spin coupled to 2 inequivalent neighbouring nuclear spins. Figure S22 shows that, based on this model, our data is not sufficient to distinguish between the cases where the central electronic spin is coupled to one B and one N, two B or two N.

Figure S23 shows the calculated ODMR spectra for alternative configurations comprising a defect coupled to $n$ equivalent nitrogen (top row) or boron (bottom row) nuclear spins, with $n$ =1-5. Amongst these configurations, we could not identify a situation that reproduces our data well. We hold that our results are more compatible with a complex carbon-cluster-related defect with low-symmetry, where the nearest neighbor nuclear spins would not be of a single defect species.

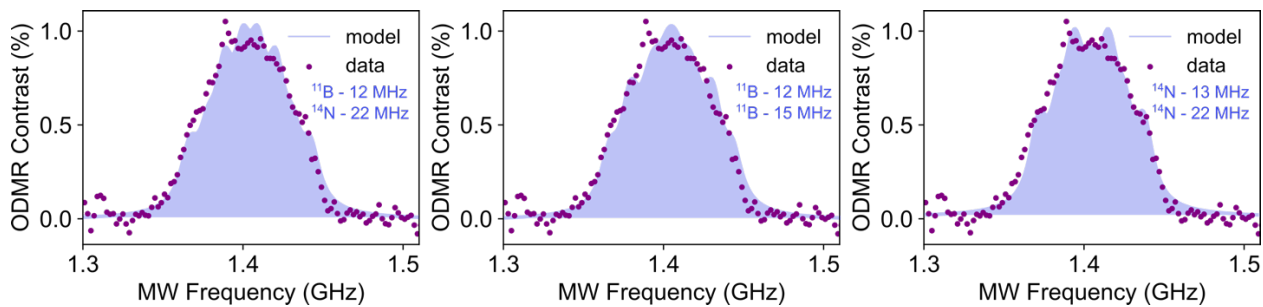

**Figure S22: Modeled ODMR lineshape for three models considering coupling between the central electronic spin and two inequivalent neighboring nuclei.** Within the resolution of our data, we cannot distinguish between the situations where the central spin is coupled to two inequivalent nuclei of different types (one boron and one nitrogen, left panel) or of the same type (two boron, central panel, or two nitrogen, right panel). Nonetheless, this is the only model that captures the ODMR lineshape in the presence of a magnetic field parallel to the symmetry axis of the defect.

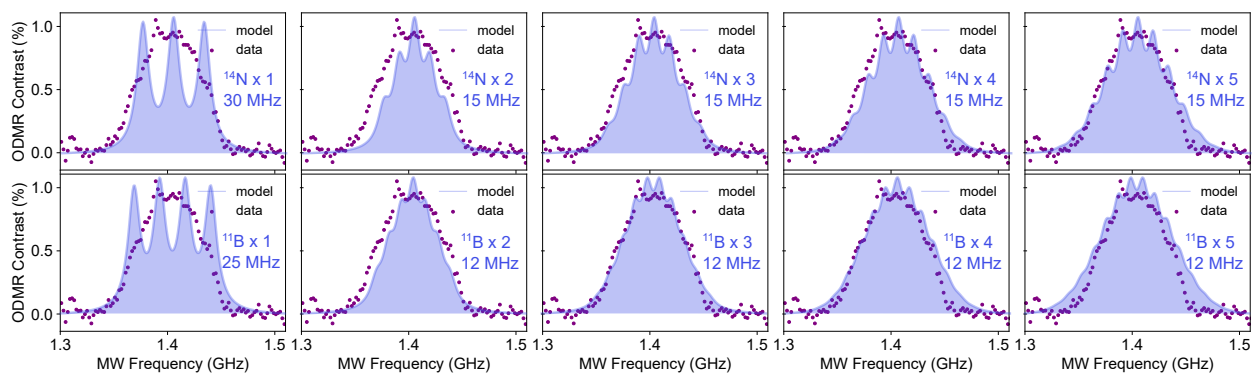

**Figure S23: Modeled ODMR lineshape for various nearest-nuclear spin configurations for an electronic spin-1 system.** The top row shows the simulation when the electronic spin is coupled to $n$ equivalent nitrogen-14 (top row) or boron-11 (bottom row) nuclei, with hyperfine coupling on the order of 15 MHz. From left to right, $n$ runs from 1 to 5. For these simulations, we assume a 35 mT magnetic field along the $z$-axis of the defect electronic spin.

# 6) References

[1] Binder, J. M., et al. Qudi: A modular python suite for experimental control and data processing. *Original Sotfware Publication*, **6**, 85-90 (2017).

[2] Berthel, M., et al. Photophysics of single nitrogen-vacancy centers in diamond nanocrystals. *Phys. Rev. B*, **91**, 3, 035308, (2015).

[3] Tetienne, J-P., et al. Magnetic-field-dependent photodynamics of single NV defects in diamond: an application to qualitative all-optical magnetic imaging, *New J. Phys.*, **14**, 103033 (2012).

[4] Babar, R., et al. Quantum sensor in a single layer van der Waals material. *arXiv 2111.09589* (2021).

[5] Oniuzhuk, M., et al. Probing the coherence of solid-state qubits at avoided crossings, *PRX Quantum*, **2**, 010311 (2021).

[6] Bayliss, S. L., et al. Enhancing Spin Coherence in Optically Addressable Molecular Qubits through Host-Matrix Control, *Phys. Rev. X.* **12**, 031028 (2022).